\documentclass[twocolumn]{aastex63}
\usepackage[utf8]{inputenc}
\usepackage{booktabs}
\usepackage{amsmath}
\newcommand{\angstrom}{\textup{\AA}}
\usepackage{amssymb}
\usepackage{physics}

\shorttitle{Hyperfine Structure in  \ion{Co}{2}}
\shortauthors{Ding \& Pickering}
\graphicspath{{./}{figures/}}
\begin{document}

\title{Measurements of the Hyperfine Structure of Atomic Energy Levels in \ion{Co}{2}}

\correspondingauthor{Milan Ding}
\email{milan.ding15@imperial.ac.uk}

\author{Milan Ding}
\affil{Physics Dept., Imperial College London, Prince Consort Road, London SW7 2AZ, UK\\}

\author{Juliet C. Pickering}
\affil{Physics Dept., Imperial College London, Prince Consort Road, London SW7 2AZ, UK\\}




\begin{abstract}

Analysis of hyperfine structure constants of singly ionised cobalt (\ion{Co}{2}) were performed on cobalt spectra measured by Fourier transform spectrometers in the region $3000-63000$~cm$^{-1}$ ($33333-1587$~$\angstrom$). Fits to over $700$ spectral lines led to measurements of $292$ magnetic dipole hyperfine interaction $A$ constants, with values between $-32.5$~mK and $59.5$~mK ($1$~mK~$=0.001$~cm$^{-1}$). Uncertainties of $255$ A constants were between $\pm0.4$~mK and $\pm3.0$~mK, the remaining $37$ ranged up to $\pm7$~mK. The electric quadrupole hyperfine interaction $B$ constant could be estimated for only $1$ energy level. The number of \ion{Co}{2} levels with known $A$ values has now increased tenfold, improving and enabling the wider, more reliable and accurate application of \ion{Co}{2} in astronomical chemical abundance analyses.

\end{abstract}

%\keywords{atomic data --- line: profiles --- methods: laboratory: atomic}


\section{Introduction} \label{sec:intro}
Advances in the resolution of astrophysical spectra have revealed the importance of physical effects that were once thought to be insignificant, one of which is the hyperfine structure (HFS) splitting of spectral lines \citep{Kurucz1993}. The HFS of energy levels involved in a transition introduces broadening and asymmetry of the spectral line profile, and if not accounted for, inaccuracies in line position and line strength. Accurate data for HFS is therefore crucial to meaningful astrophysical chemical abundance analysis carried out, for example, in studies of stellar evolution, galactic chemical evolution and nucleosynthesis.

Recently, the best available HFS data for each iron-group element were applied by \cite{Scott2015} to estimate the most accurate solar iron-group elemental abundances. However, no spectral lines of \ion{Co}{2} were found suitable for the analysis, reminiscent of the unsuccessful synthesis of several observed solar spectral lines of \ion{Co}{2} by \cite{Biemont1978}, which appeared too broad and were unable to be accounted for without \ion{Co}{2} HFS data.

\cite{Bergemann2010} performed the first HFS measurements for $6$ magnetic dipole interaction ($A$) constants of \ion{Co}{2} using high resolution Fourier transform (FT) spectroscopy. The $A$ constants were used to test new nLTE (non-local thermodynamic equilibrium) modelling of stellar spectral lines. Their results showed that the neglect or use of incorrect $A$ values produces significantly different line profiles to observation, and hence unreliable abundance estimations. A further set of $22$ $A$ constants for the low-lying \ion{Co}{2} levels were first measured by \cite{Lawler2018}. Applying these in the abundance analysis of the metal-poor star HD~84937 resulted in a change of about $-0.1$~dex in the Co/Fe ratio, compared to previous analyses without laboratory measured A constants by \cite{Lawler2015} and \cite{Sneden2016}.

Without accurate characterisation of HFS of energy levels involved in atomic transitions used by astronomers, accurate abundance analyses may be difficult or impossible. There is a need for all relevant HFS data of energy levels, so that the analyses of modern high resolution astrophysical spectra are not limited by the quality and quantity of atomic data available.

This paper presents measurements of HFS $A$ constants using FT spectra of cobalt. New HFS $A$ constants of \ion{Co}{2} are reported for $292$ levels, of which $264$ are measured for the first time.

\section{Experimental Details}
Fourier transform spectroscopy is the ideal technique for the determination of a large number of HFS A constants, since its wide spectral range and high resolution enables the analysis of many hundreds of spectral lines from the infrared (IR) to ultraviolet (UV) regions. 

The spectra used in this analysis were a part of the extensive measurement of \ion{Co}{1} and \ion{Co}{2} spectra in the region $3000-63000$~cm$^{-1}$ ($33333-1587$~$\angstrom$), which contributed to the comprehensive revision and identification of lines and energy levels of \ion{Co}{1} by \cite{PickeringThorne1996} and \ion{Co}{2} by \cite{Pickering1998}, \cite{Pickering1998a} and  \cite{Pickering1998b}. \cite{Pickering1996} also determined, at the time, $208$ previously unknown \ion{Co}{1} $A$ constants using the same set of laboratory spectra.

Spectra in the visible-UV region were measured with the f/25 UV FT spectrometer at Imperial College. Infrared spectra were recorded with the 1 m f/55 FT spectrometer at the National Solar Observatory in Arizona. Spectral resolutions were $0.05$~cm$^{-1}$ ($0.003$~$\angstrom$ at $2500$~$\angstrom$) in the UV, $0.036$~cm$^{-1}$ ($0.006$~$\angstrom$ at $4000$~$\angstrom$) in the visible and $0.0139$~cm$^{-1}$ ($0.028$~$\angstrom$ at $10000$~$\angstrom$) in the IR, chosen such that the resolution of individual fine structure lines was limited by Doppler width, as opposed to the instrumental profile. The light source was a water-cooled hollow cathode lamp, run at currents of $600-700$~mA using a pure Co cathode, and either Ar (at $1$~mbar) or Ne (at $2-3$~mbar) were used as the carrier gas for each spectrum. Full experimental details are given in \cite{Pickering1998}.

The wavenumbers and energy level values of \ion{Co}{2} reported in \cite{Pickering1998}, \cite{Pickering1998a} and \cite{Pickering1998b} were wavenumber calibrated using \ion{Ar}{2} reference lines measured by \citet{Norlen1973}. As recommended by \citet{Nave2011}, the energy level and centre of gravity line wavenumbers have been increased in this work by $6.7$ parts in $10^8$ to place them on the revised \ion{Ar}{2} wavenumber scale determined by \citet{Whaling1995}.

\section{Theory of Observed Line Profiles}\label{theory}
\begin{figure}
    \centering
    \includegraphics[width = \linewidth]{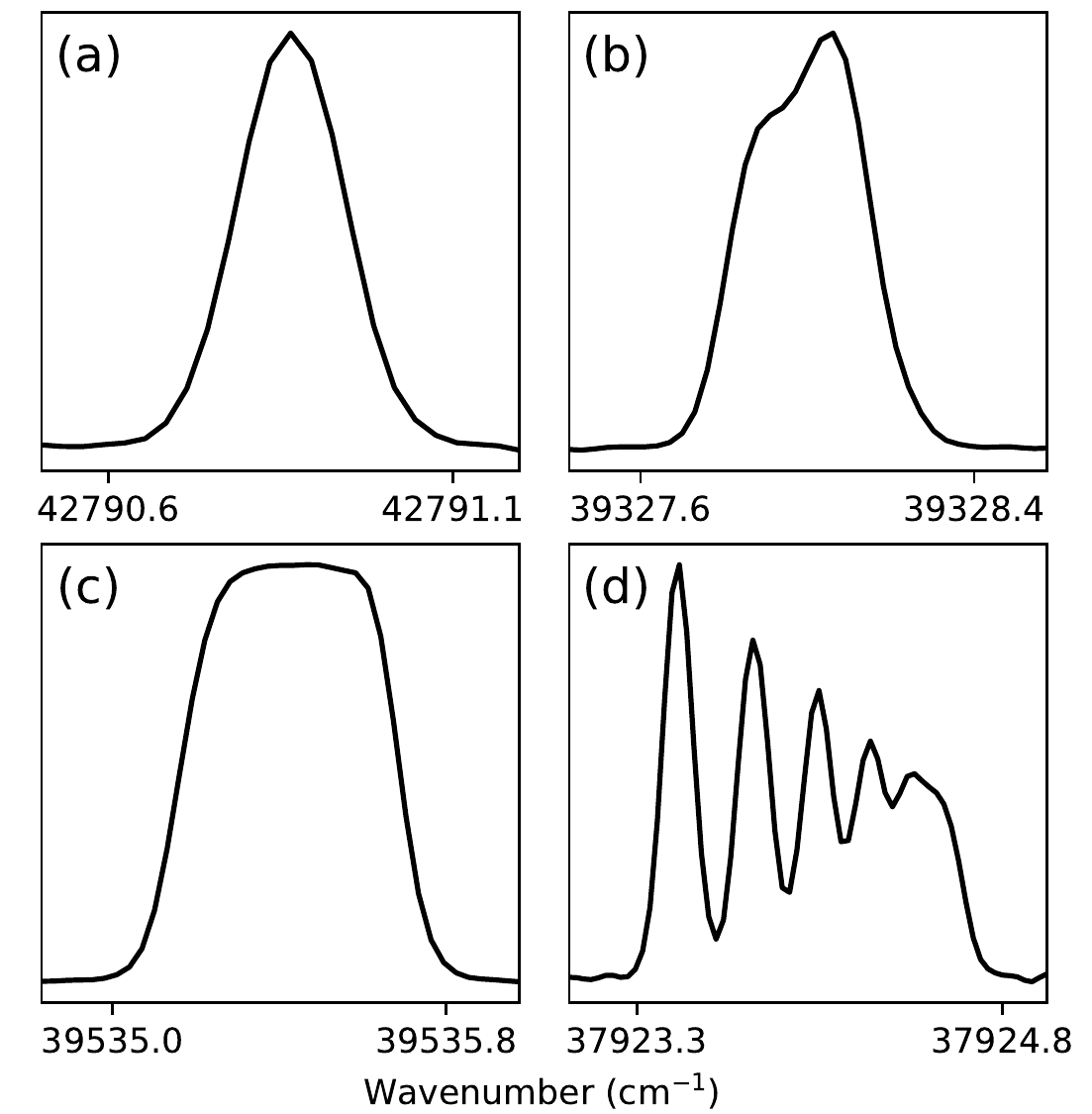}
    \caption{Normalised intensity profiles of $4$ observed fine structure lines of \ion{Co}{2}. All other \ion{Co}{2} line profiles showed at least one of these $4$ characteristics: (a) symmetric, (b) asymmetric, (c) broad or (d) with clear, pronounced spectral features. Note that the wavenumber axes differ in scale.}
    \label{figure1}
\end{figure}

Fine structure lines from the FT emission spectra of \ion{Co}{2} displayed a range of profiles, and typical examples are shown in figure \ref{figure1}. There is only one stable isotope of cobalt, $^{59}$Co with nuclear spin $7/2$, and its HFS constants were measured by fitting model line profiles to the observed spectral lines.

\subsection{Hyperfine Structure}
For an atom with nucleus of spin quantum number $I$, individual fine structure atomic energy levels are split due to the interaction between the associated nuclear and total electronic magnetic dipole moments. Assuming no other perturbations, the energies of the hyperfine levels are shifted from the fine structure energy according to \citep{Kopfermann1958}:
\begin{equation}
    \Delta W_{FJ}= \frac{1}{2}AK + B \frac{(3/4)K(K+1)-J(J+1)I(I+1)}{2I(2I-1)J(2J-1)},
\end{equation}
\noindent where $\Delta W_{FJ}$ is the change in energy from the fine structure level with angular momentum $J$ to its corresponding HFS level with total atomic angular momentum $F$.
\begin{equation*}
    F = I+J;I+J-1;\dots;\abs{I-J},
\end{equation*}
and $K$ is defined as
\begin{equation*}
    K = F(F+1)-J(J+1)-I(I+1).
\end{equation*}
$A$ and $B$ are the magnetic dipole and electric quadrupole hyperfine interaction constants respectively. Within a fine structure transition (HFS pattern), the relative intensity of a hyperfine component transition, $R_{JFJ'F'}$, between the hyperfine levels with quantum numbers $JF$ and $J'F'$ is given by
\begin{equation}
    R_{JFJ'F'}=\frac{(2F+1)(2F'+1)}{(2I+1)}
    \begin{Bmatrix}
    J & F & I \\
    F' & J' & 1
    \end{Bmatrix}^2,
\end{equation}
where the squared factor is a Wigner $6\text{-}j$ symbol, which includes the selection rules: $\Delta F = 0,\:\pm 1$ but not $F=0\leftrightarrow F'=0$  \citep{Cowan1981, Wahlgren1995}.

\subsection{Spectral Line Profile} \label{lineprof}
For a transition between two fine structure levels, the hyperfine $A$ and $B$ constants of the levels determine the wavenumbers of all allowed hyperfine component transitions, relative to the centre of gravity wavenumber of the transition. 

\begin{figure}
    \centering
    \includegraphics[width = \linewidth]{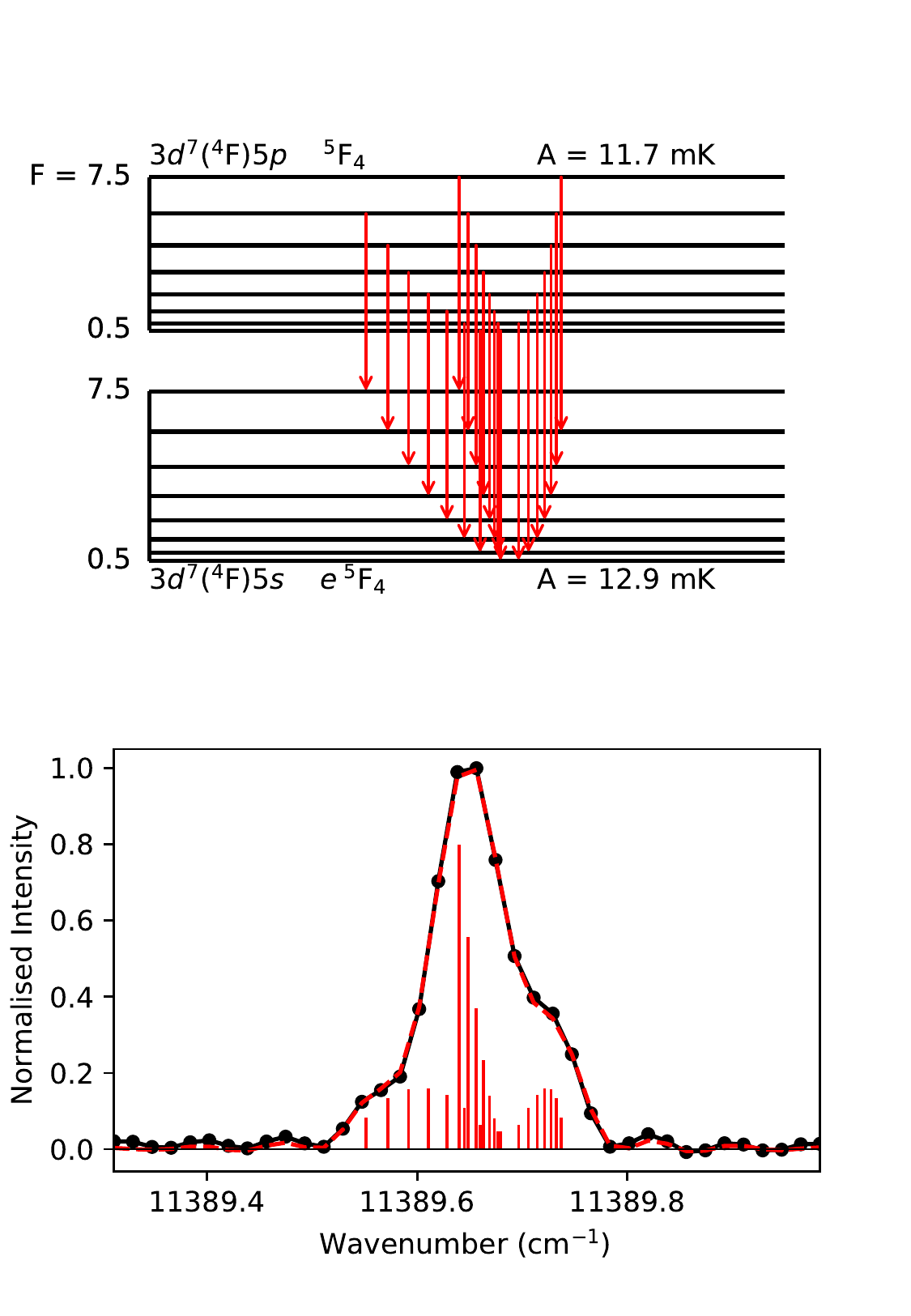}
    \caption{Observed (black) and fitted (red dashed) line profile of the $11389.710$~cm$^{-1}$ transition, with a signal-to-noise ratio (SNR) of $160$. Relative intensities of HFS components (red) and transition diagram are shown.}
    \label{figure2}
\end{figure}

Without any other forms of energy level splitting, the individual hyperfine component transitions observed from hollow cathode discharge lamps are emission spectral lines of a plasma, described by the Voigt profile \cite{Thorne1999}. Therefore, fine structure lines in the FT spectra of \ion{Co}{2} were observed as blends of several Voigt profiles, defined by relative intensities and hyperfine interaction constants $A$ and $B$. An example composition of an observed asymmetric line profile is shown in figure \ref{figure2}.

A Voigt spectral line profile results from the dominant Doppler and pressure broadening effects in the gas, which are characterised by the Gaussian and Lorentzian width components respectively in the Voigt Profile.

\subsection{Instrumental Line Profile}
An observed line profile from an FT spectrometer is the convolution between the true profile and its instrumental profile. The instrumental profile is mostly dependent on source-aperture geometry and interferogram apodisation procedures, but its functional form can be difficult to fully specify. For the purposes of fitting observed lines, apodising the FT of the model line profile with a box funcion sufficiently approximates any ringing artifacts and distortion due to the instrumental profile \citep{Davis2001}. This is only usually necessary in modelling line profiles observed under lower resolving powers, which is readily seen from figure \ref{figure2}, where the peaks without any component transitions on the sides were ringing artifacts significantly above the noise level.

\section{Hyperfine Structure Analysis}

\subsection{Least Squares Fitting}
As indicated in section \ref{theory}, a number of parameters are required to model the spectral line of a transition between fine structure energy levels of \ion{Co}{2}. The $J$ quantum numbers for both levels of each transition were known prior to this work, hence the following 8 parameters required fitting in order to determine the HFS constants:
\begin{enumerate}
    \item $A$ and $B$ HFS constants of the two fine structure levels of the transition ($4$ parameters).
    \item Component line Gaussian width $G_w$.
    \item Component line Lorentzian width $L_w$.
    \item Total intensity.
    \item Centre of gravity wavenumber.
\end{enumerate}

The \textit{hfs\_fit} program written during this work was the central tool of the analysis. Model lines were produced from the parameters, and the square of the residuals was minimised by optimising the parameters using a simulated annealing algorithm. Optimisations were classified as complete when the residuals were of the order of the noise.

\textit{hfs\_fit} is capable of inspection of parameter dependence; comparison between model and observed profiles are shown visually under parameter variation. Any parameter can be held constant for an optimisation. The probability distributions for parameter values at each iteration within an optimisation are Gaussian, their widths can be modified, as necessary. These features support the algorithm in reaching the global minimum by allowing supervision and the choice of suitable initial parameter values.

\subsection{Parameter Constraints}
Of the 8 parameters computationally optimised, line profiles were predominantly most sensitive to the $A$ constants and $G_w$. Centre of gravity wavenumber and total intensity only affect position and area of the profiles respectively, they do not affect the profile shape. Accurate initial values of centre of gravity wavenumber and total intensity were calculated from the wavenumber and area of the observed line data, or specified by inspection. $L_w$ was about $10$~mK and approximately constant within each spectrum, in comparison, $G_w$ varied from $20$~mK to $100$~mK from the IR to visible and $120$~mK to $220$~mK across the UV, where the majority of the stronger \ion{Co}{2} lines lie. Fits to all profiles showed much smaller sensitivity to $B$ constants compared to that of the $A$ constants, values of $B$ are also in general one order smaller than the $A$ constants (e.g. see \cite{Pickering1996}), so the initial values for $B$ in the fitting process were always set to $0$~mK.
\begin{figure}[t]
    \centering
    \includegraphics[width = \linewidth]{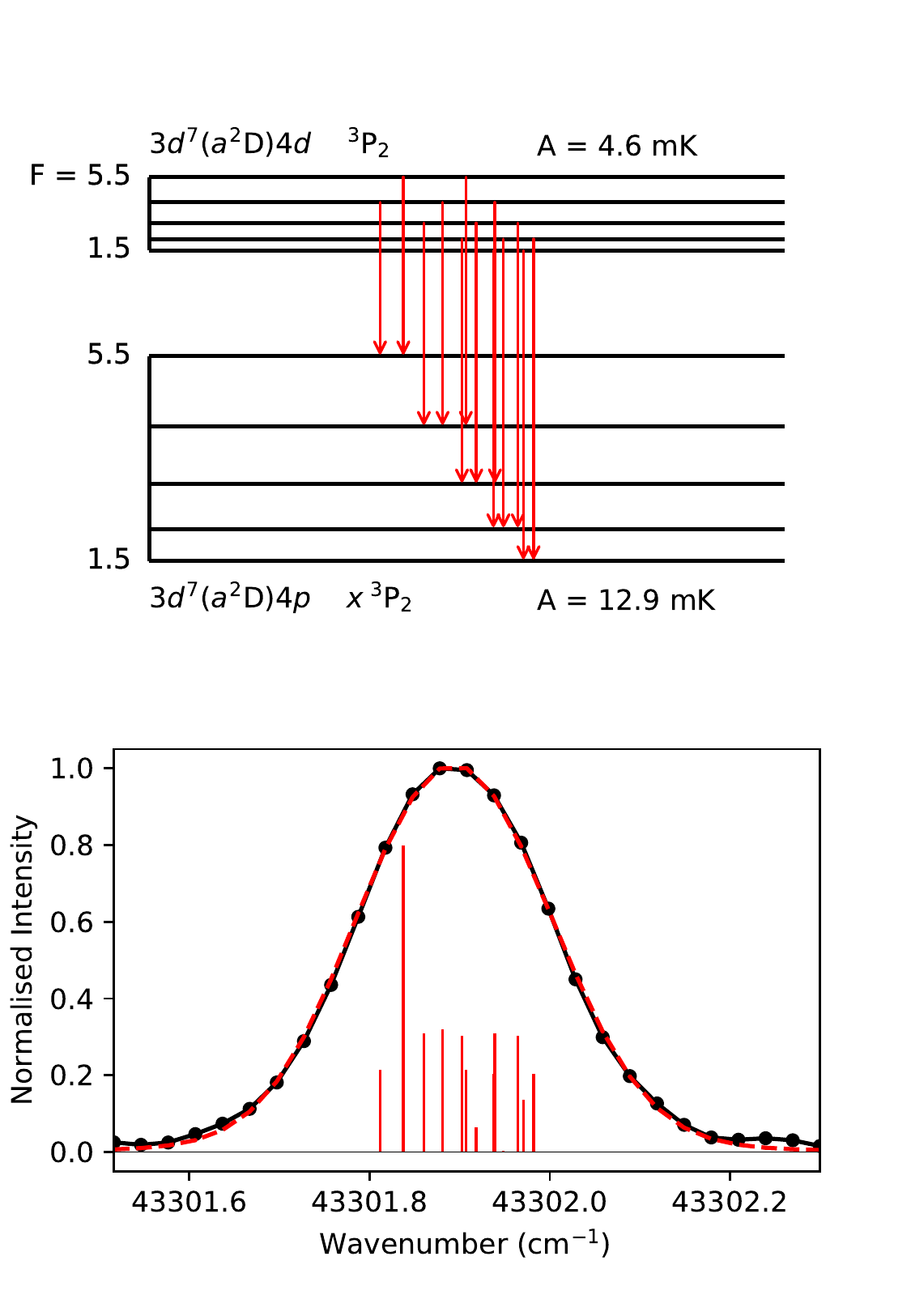}
    \caption{Observed (black) and fitted (red dashed) line profile of the $43301.944$~cm$^{-1}$ transition, SNR$\:=80$. Relative intensities of HFS components (red) and transition diagram are shown. \label{figure3}}
\end{figure}
\begin{figure}[t]
    \centering
    \includegraphics[width = \linewidth]{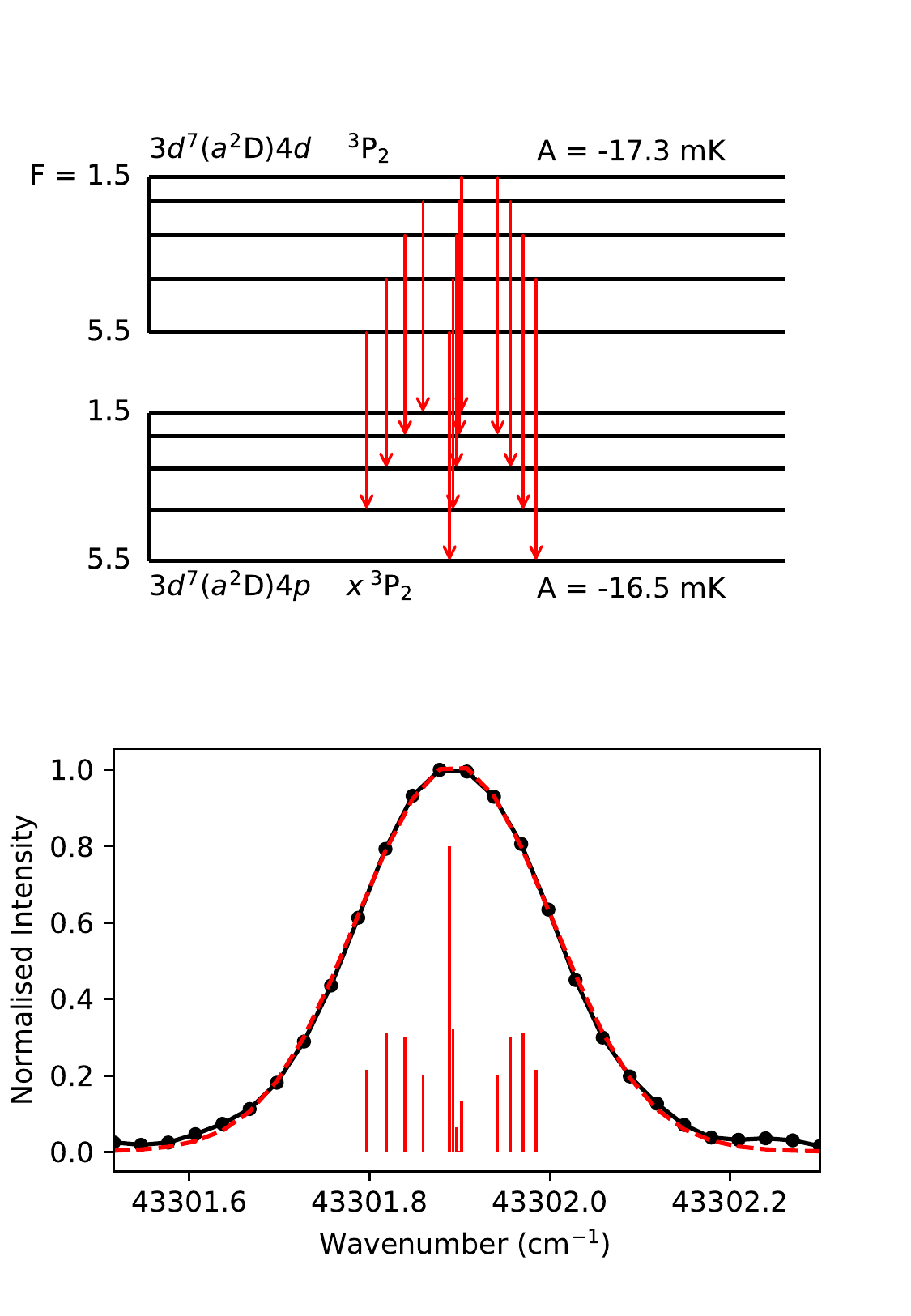}
    \caption{Fit of the line from figure \ref{figure3} using a different pair of $A$ constants and a very similar $G_w$. These $A$ constants fit the line, but do not agree with values indicated by other transitions involving either of the two levels. \label{figure4}}
\end{figure}

Even with the above stated constraints, optimisations were not possible given arbitrary initial values of $A$ constants and $G_w$. Furthermore, without any constraint on these $3$ parameters, profiles such as the ones shown in figure \ref{figure1} (a), (b) and (c) would give a wide range of optimal values, which limited the accuracy of $A$ constant measurements. In the worst cases, the optimal $A$ constants could vary by more than $20$~mK, and an example is shown in figure \ref{figure3}. For this type of line profile, a redistribution of components (i.e. having different $A$ constants to those shown), combined with slight changes in their width $G_w$ and total intensity could produce the same observed profile, this is seen from figure \ref{figure4}.
\subsection{Analysis Strategies}
When performing HFS analysis on FT spectra with line profiles summarised in figure \ref{figure1}, it is crucial to have further constraints on the $A$ constants and $G_w$, and these can indeed be obtained as the analysis progresses.

Knowing one of the two $A$ constants during the analysis of a line profile greatly reduces the range of possible optimal values for the other $A$ and $G_w$. When the unknown $A$ constant is accurately determined this way, it could then serve as a new \textit{reference} $A$ constant for other transitions involving its corresponding fine structure energy level. This process could be iterated from a set of known $A$ constants used as references to increasingly include a larger number of energy levels, and their $A$ constants could thus be sequentially determined.

Due to the iterative nature of the analysis method, values and uncertainties of subsequently determined $A$ constants depend on the initial set of reference $A$ constants. Highly accurate measurements, e.g. by laser or Fabry P\'{e}rot techniques, of $A$ and $B$ constants are not available for \ion{Co}{2}. This \ion{Co}{2} HFS analysis therefore began with the assumption of no previously known $A$ or $B$ constants.

\begin{figure}[t]
    \centering
    \includegraphics[width = \linewidth]{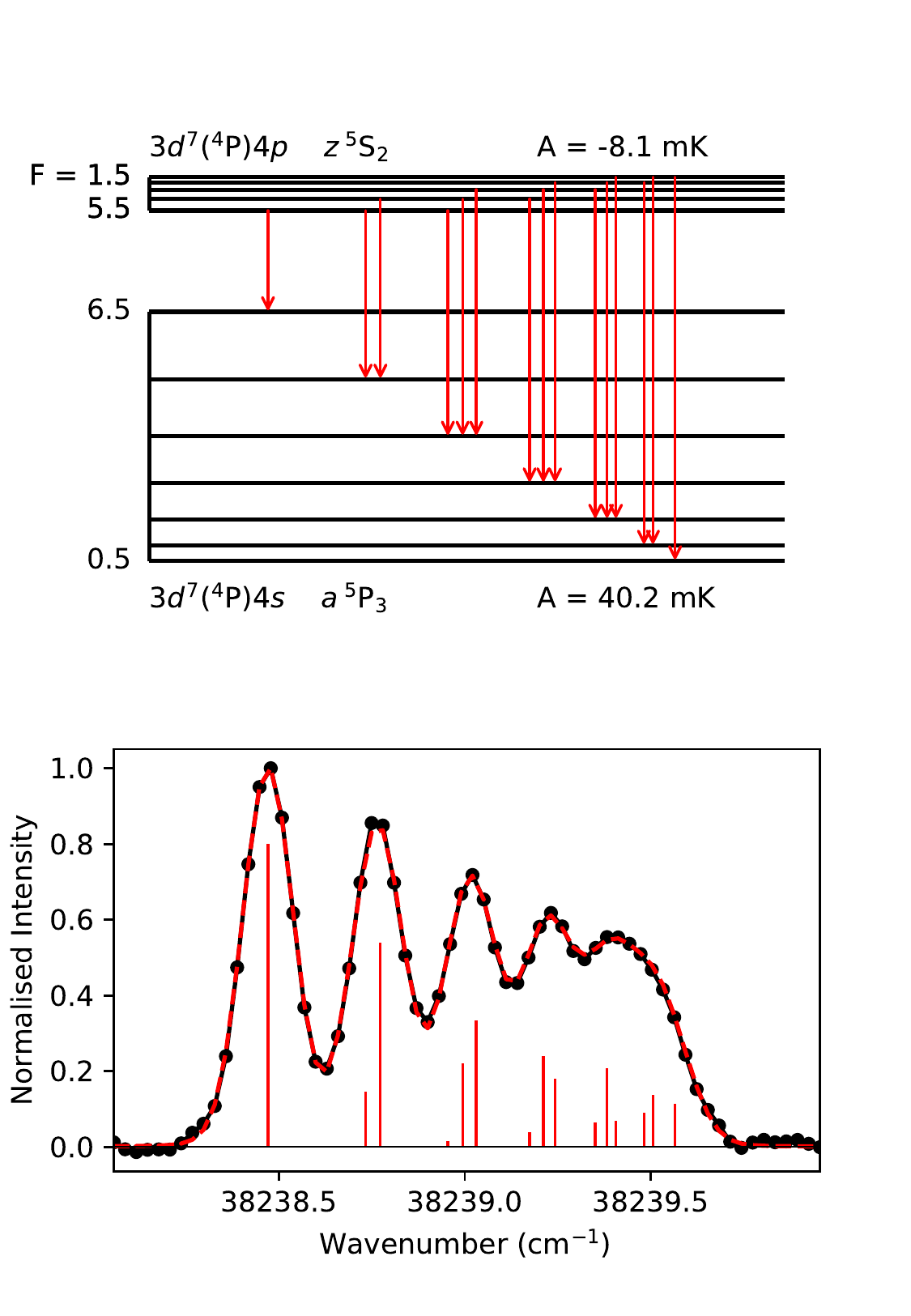}
    \caption{Observed (black) and fitted (red dashed) line profile of the $38238.969$~cm$^{-1}$ transition, SNR$\:=70$. Relative intensities of HFS components (red) and transition diagram are shown.}
    \label{figure5}
\end{figure}

Profiles offering the most constraints on $A$ constants and $G_w$ were selected for the set of initial reference $A$ constants. They showed pronounced spectral features and their $A$ constants could be determined with uncertainties ranging from $\pm 1$~mK up to $\pm 3$~mK, without any prior information on the $A$ values or $G_w$. Figure \ref{figure5} is an example of such profiles. Here, the highest intensity hyperfine component transition was clearly separated from others, which also allowed accurate determination of $G_w$. A total of $18$ lines with similar profiles were chosen in the $36000$~cm$^{-1}$ to $42000$~cm$^{-1}$ spectral range from two UV spectra containing the largest number of \ion{Co}{2} lines. The $A$ constants of the $18$ lines were then determined by fitting and checking for agreement amongst other lines that involved them. Any line involving a level with $J=0$ naturally possesses tighter constraints on the values of $A$ and $G_w$, since the $A$ of the $J=0$ level is zero. However, their profiles did not show pronounced spectral features and their unknown $G_w$ hindered the determination of their $A$ constants at this stage.
\begin{figure}[b]
    \centering
    \includegraphics[width = \linewidth]{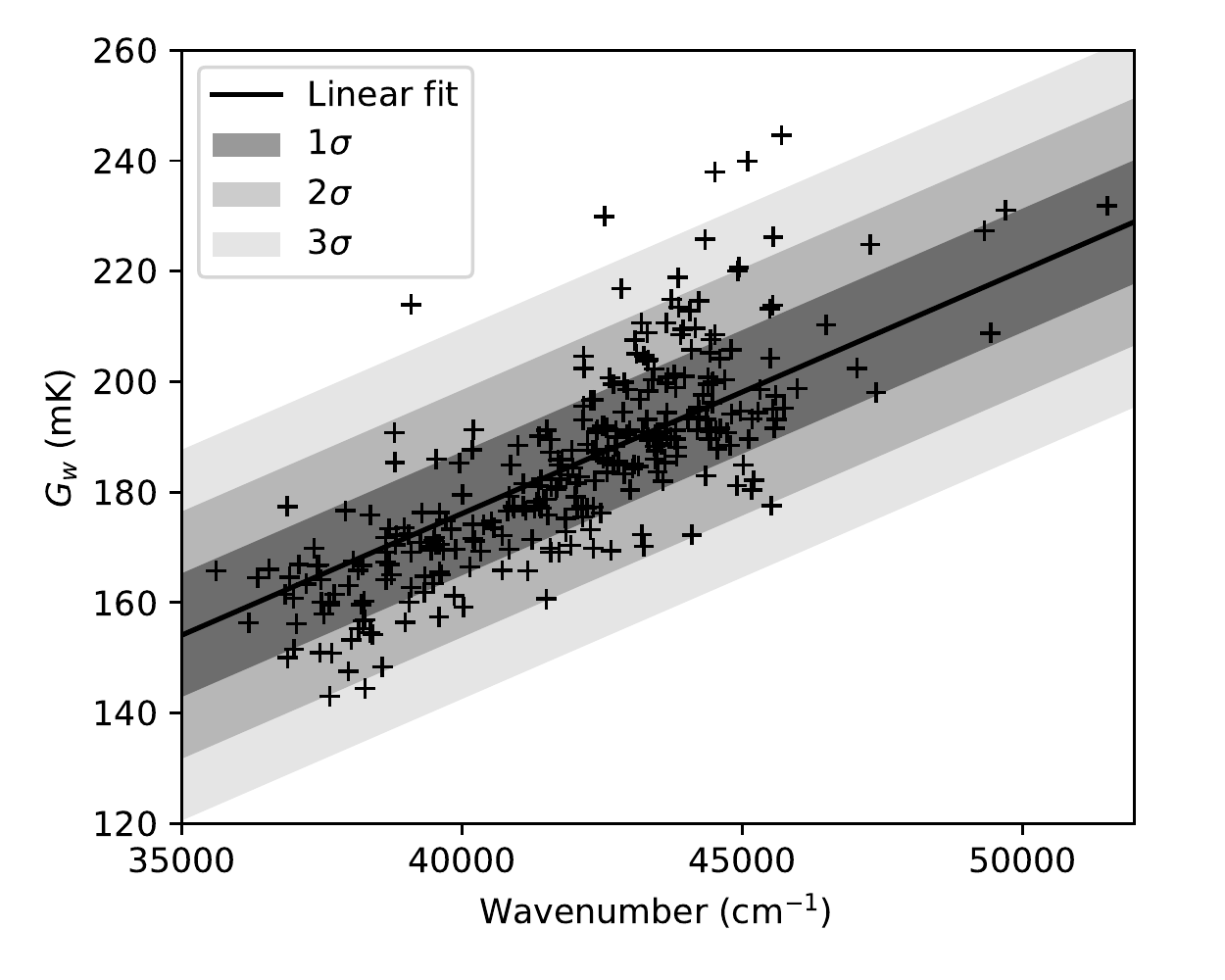}
    \caption{$G_w$ against wavenumber for $307$ lines fitted in a UV spectrum. Linear fit of the Doppler width relation and standard deviations of residuals are shown.}
    \label{figure6}
\end{figure}

This initial analysis not only produced a set of reference $A$ constants, but also covered a broad spectral range within the two UV spectra, which provided an essential constraint on $G_w$. For a Maxwellian gas, the relationship between $G_w$ and wavenumber is linear. The $G_w$ obtained from all fits in a particular UV spectrum are plotted against their wavenumber in figure \ref{figure6}. This relationship had a Pearson correlation coefficient of $0.8$ for all spectra analysed, which was also observed by \cite{Pickering1996} in the analysis of \ion{Co}{1} HFS. When combined with the strategy of using reference $A$ constants, the unknown $A$ constant of a fit became the only unknown parameter. This enabled a large number of further fits, which were then typically for symmetric profiles (e.g. figure \ref{figure3}) or lower SNR profiles showing slight asymmetry. The most significant deviations, such as the five fits with $G_w$ more than $3\sigma$ above the expectation seen in figure \ref{figure6}, were suspected to be due to self-absorption.

\subsection{Mean Value and Uncertainty Estimation of HFS Constants}
To determine an $A$ constant and later use it as a reference to go on to find other $A$ constants, lines of its associated fine structure level must show agreement across multiple spectra. In fulfilling this criterion for each $A$ constant, any effects from self-absorption or blending with other lines could become apparent. When profiles failed to be fitted using expected $G_w$ or $A$ constants due to these effects, they were omitted from the estimation of associated HFS constants, this is further discussed in section \ref{self_a}.

When an $A$ constant was measured from multiple profiles observed in multiple spectra, the weighted sample mean and standard deviation were estimated. The weighting was determined by SNR and the clarity of spectral features, where the latter was quantified by the derivative of the normalised model line profile. Specifically, the unnormalised weights $w_i$ were calculated as
\begin{equation}
    w_i = \text{SNR}\times \bigg(\sum_n\abs{\frac{I_{n + 1} - I_n}{\Delta\sigma}}\bigg)^3,
\end{equation}
where $I_n$ is normalised intensity of the model line profile and $n$ is the index in the wavenumber axis. The wavenumber axis of each line profile was scaled to be from $0$ to $1$, representing the spectral range of the line profile. $\Delta\sigma$ is the scaled resolution, and so profiles with less data points were also weighted less. The power of $3$ on the summation places more weight on spectral features rather than SNR, because profiles with more asymmetry and features constrained the fitted $A$ constants and $G_w$ better. SNR was weighted less because it did not affect line profile unless it was very low and in most cases such lines could not be used in the analysis. If the SNRs of the profiles shown in figure \ref{figure1} were the same, (d) would be weighted greatest with $0.97$, which is expected as this profile constrains $A$ and $G_w$ far better than other examples in the figure. Even if profiles (a), (b) and (c) were to have $10$ times the SNR of (d), the weights from (a) to (d) would be $0.02$, $0.11$, $0.11$ and $0.76$ respectively, still in favour of profile (d). The electric quadrupole $B$ constants were also estimated using this weighted method.

Most lines did not show pronounced spectral features (e.g. figures \ref{figure1} (a), (b), (c), \ref{figure2} and \ref{figure3}), and so their $A$ constants were less well constrained. Furthermore, many levels had only such profiles and less than $5$ lines suitable for fitting. The variability of optimal $A$ values from the fits was therefore the indicator for uncertainty, and so the weighted estimation was then unused. Instead, the estimation of the $A$ constant was done on a worst-fit basis - fits were carried out at all mean values and also at extremes of possible values of the reference $A$ constant and $G_w$ (from figure \ref{figure6}) based on their uncertainties. The $B$ constants could not be estimated this way, due to profile insensitivity to $B$.

Even with the $G_w$ and reference $A$ known, there were occasional two-value ambiguities for the $A$ constant of interest. This problem was encountered by \cite{Lawler2018} and was similarly solved by checking agreement amongst multiple lines of the level with the $A$ constant.

\subsection{Self-absorption and Blends}\label{self_a}
Spectral lines affected by self-absorption can appear broadened, flattened or self-reversed \cite{Cowan1948}. Effects of self-absorption in a fine structure line were indicated by comparing its observed profiles in spectra with varying lamp conditions, particularly those with differing carrier gases (either Ar or Ne), while also examining $G_w$ trends of each spectrum and the associated $A$ constants determined from other lines. Blends could be similarly identified.

Self-absorption was modelled using a parameter $S$ that incorporated the absorption coefficient of the transition, adopted from the DECOMP program by \cite{Brault1989}:
\begin{equation}\label{saturation}
    I(\sigma) = I_0(\sigma) \: e^{-SI_0(\sigma)},
\end{equation}
\noindent where $I(\sigma)$ and $I_0(\sigma)$ are the line profiles after and prior to self-absorption respectively. Using one $S$ parameter to account for self-absorption in profiles with anomalous $G_w$ did not shift $G_w$ down to expected values. Introducing an $S$ parameter to each hyperfine component transition produced the expected $G_w$ and $A$ constants, but these fits had too many parameters for results to be meaningful. For blends, unknown HFS constants and line intensities introduce the same issue. Therefore, profiles with anomalous $G_w$ and blends were omitted from the estimation of associated $A$ constants.

\section{Results}
\subsection{Magnetic Dipole Interaction A Constants}
All measured magnetic dipole interaction constants, $A$, are reported in Tables \ref{table1} and \ref{table2} for even and odd levels respectively. The configuration, term, $J$ value and energy of each level are listed in the first $4$ columns, which were reported in \cite{Pickering1998}. The 5th and 6th column list the A constants and their uncertainties in mK ($1$~mK = $0.001$~cm$^{-1}$). The number of line profiles used to determine $A$ values and their uncertainties, $N$, is indicated in the seventh column. Previous measurements are given in the last column where available.

Agreement was found within joint uncertainties between previously published values by \cite{Bergemann2010} and \cite{Lawler2018} and the new $A$ constants of this work. The uncertainties for each $A$ constant have been carefully estimated allowing its contribution to any resulting astrophysical chemical abundance determination uncertainties to be calculated. In the vast majority of cases, the small uncertainties of the $A$ constants are expected to have a negligible effect on the resulting abundance uncertainties, given typical observational uncertainties.

As expected, the $A$ constants were typically larger for levels involving an unpaired $s$ valence electron, and they became smaller as the principal and angular momentum quantum numbers of the valence electron increased. One $A$ constant was found for the $5d$ levels, which was for the $3d^7(^4\text{F})5d\:\:^5\text{G}_2$ level. No HFS analysis was carried out for the $5g$ levels. Other $5d$ levels and all $5g$ levels are excluded from the tables.
\subsection{Electric Quadrupole Interaction B Constants}
The electric quadrupole interaction constants, $B$, were unable to be estimated accurately. Line profiles were mostly sensitive to the difference in the $B$ constants of the profiles, rather than in the particular B values themselves. The resulting uncertainty from weighted calculation was always similar to, if not larger than the estimated means. Hence, they are not reported in the table.

However, lines involving the $3d^7(^4\text{P})4s\:\:a\:^5\text{P}_2$ level consistently indicated a positive value for its $B$ constant. The $14$ profiles investigated for this level gave estimates of $B$ as $5\pm4$~mK.

\section{Conclusion}
This work reports the measurement of $292$ magnetic dipole interaction $A$ constants of \ion{Co}{2}, of which $264$ were previously unknown. It is shown that with FT spectra ranging from the IR to UV, the existence of a set of lines exhibiting pronounced HFS splitting enables a large number of $A$ constants of an atom to be determined, even when almost all lines have no observable individual HFS component transitions. One $B$ constant is reported. The $B$ constants were impossible to determine accurately, similar to other HFS analysis using FT spectra, e.g. \cite{Pickering1996}, \cite{Palmeri1995}, \cite{Townley2016} and \cite{Lawler2018}.

The HFS of a large fraction of all classified, observed transitions of \ion{Co}{2} is now characterised. Wider, more accurate and reliable application of \ion{Co}{2} in astronomical chemical abundance analyses is now enabled.

\acknowledgements{This work was supported by the STFC (UK). The authors thank Christian Clear for helpful comments on the manuscript. The authors also thank Gillian Nave for her advice during the initial stages of this project.}

\software{astropy \citep{Robitaille2013}, matplotlib \citep{Hunter2007}, numpy \citep{Walt2011}, pandas \citep{McKinney2010}}

%\input{tab3}
%\input{tab4}
%\section*{}
%\clearpage

\bibliography{bibliography}{}
\bibliographystyle{aasjournal}

\begin{deluxetable*}{lllrrrll}
\tablecolumns{8}
\tablecaption{Hyperfine Structure Magnetic Interaction A Constants of the Even-parity Energy Levels of \ion{Co}{2}}
\label{table1}
%\tabletypesize{\scriptsize}
\tablehead{\colhead{Configuration} & \colhead{Term} & \colhead{J} & \colhead{Energy} & \colhead{A} & \colhead{Unc.} & \colhead{N} & \colhead{Previous Work} \\
                                  &  &  &  \colhead{(cm$^{-1}$)} & \colhead{(mK)} & \colhead{(mK)} & & \colhead{(mK)}
                              }

\startdata
                       $3d^8$ &      $a\:^3$F &  4 &               0.000 &    13 &         5 &   2 &                                       \\
                              &               &  3 &             950.324 &    17 &         5 &   2 &                                       \\
                              &               &  2 &            1597.197 &    33 &         4 &   1 &                                       \\
                              &&&&&&&\\
         $3d^7(^4\text{F})4s$ &      $a\:^5$F &  5 &            3350.494 &    34.4 &         1.0 &   7 &                     $33.8\pm0.8$$^a$ \\
                              &               &  4 &            4028.988 &    29.5 &         1.8 &   9 &                     $29.9\pm0.8$$^a$ \\
                              &               &  3 &            4560.789 &    25.4 &         2.2 &   6 &                     $25.0\pm0.8$$^a$ \\
                              &               &  2 &            4950.062 &    17.1 &         2.0 &   8 &                     $17.6\pm0.8$$^a$ \\
                              &               &  1 &            5204.698 &    -8.4 &         1.2 &   5 &                     $-9.2\pm1.0$$^a$ \\
\enddata
\tablecomments{Energy level values were taken from \cite{Pickering1998}, \cite{Pickering1998a} and \cite{Pickering1998b} with correction of $6.7$ parts in $10^8$ applied to place the levels on the revised wavenumber scale as recommended by \cite{Nave2011}. N is the number of observed profiles used to determine the value and uncertainty of the A constants. $^a$\cite{Lawler2018}. Only a portion of this table is shown here to show its form and content. The full machine-readable table is available in the online version of this paper.}
\end{deluxetable*}

\begin{deluxetable*}{lllrrrll}
\tablecolumns{8}
\newpage
\tablecaption{Hyperfine Structure Magnetic Interaction A Constants of the Odd-parity Energy Levels of \ion{Co}{2} \label{table2}}
%\tabletypesize{\scriptsize}
\tablehead{
                \colhead{Configuration} &          \colhead{Term} &  \colhead{J} &  \colhead{Energy}      &  \colhead{A}   &  \colhead{Unc.} &   \colhead{N} &                    \colhead{Previous Work} \\
                              &               &    &  \colhead{(cm$^{-1}$)} & \colhead{(mK)} & \colhead{(mK)}   &     & \colhead{(mK)}
                              }

\startdata
         $3d^7(^4\text{F})4p$ &      $z\:^5$F &  5 &           45197.711 &    12.1 &         1.5 &   5 &                    $10.9\pm0.8$$^a$ \\
                              &               &  4 &           45378.754 &     8.5 &         1.5 &   6 &                     $9.3\pm0.8$$^a$ \\
                              &               &  3 &           45972.036 &    13.8 &         3.1 &   4 &                    $11.5\pm0.8$$^a$ \\
                              &               &  2 &           46452.700 &    18.6 &         1.0 &   4 &                                      \\
                              &               &  1 &           46786.409 &    45.6 &         1.0 &   1 &                                      \\
                              &&&&&&&\\
         $3d^7(^4\text{F})4p$ &      $z\:^5$D &  4 &           46320.832 &     8.8 &         1.7 &   3 &   $8.8\pm0.8$$^a$, $9\pm2$$^b$ \\
                              &               &  3 &           47039.105 &     8.1 &         2.2 &   5 &  $7.8\pm0.8$$^a$, $8\pm20$$^b$\\
                              &               &  2 &           47537.365 &     8.0 &         2.6 &   3 &                     $9.0\pm0.8$$^a$ \\
                              &               &  1 &           47848.781 &     2 &         4 &   3 &   $5.5\pm0.7$$^a$, $5\pm5$$^b$ \\
                              &               &  0 &           47995.594 &     0 &         0 &     &                                      \\
\enddata
\tablecomments{Energy level values were taken from \cite{Pickering1998}, \cite{Pickering1998a} and \cite{Pickering1998b} with correction of $6.7$ parts in $10^8$ applied to place the levels on the revised wavenumber scale as recommended by \cite{Nave2011}. N is the number of observed profiles used to determine the value and uncertainty of the A constants. $^a$\cite{Lawler2018}. $^b$\cite{Bergemann2010}. Only a portion of this table is shown here to show its form and content. The full machine-readable table is available in the online version of this paper.}

\end{deluxetable*}



\end{document}